\documentclass[11pt]{article}
\usepackage{txfonts}
\usepackage{color,epsfig}
\usepackage{mathrsfs}
\usepackage{amssymb}
\usepackage{amsfonts,graphicx,psfrag}

\newtheorem{lemma}{Lemma}[section]
\newtheorem{theorem}[lemma]{Theorem}

\newtheorem{proposition}[lemma]{Proposition}
\newtheorem{definition}[lemma]{Definition}

\newtheorem{remark}[lemma]{Remark}
\let\lutzremark=\remark
\def\remark{\lutzremark\normalfont}

\newtheorem{notation}[lemma]{Notation}

\def\be{\begin{equation}}
\def\ee{\end{equation}}
\def\bea{\begin{eqnarray}}
\def\eea{\end{eqnarray}}
\def\bes{\begin{eqnarray*}}
\def\ees{\end{eqnarray*}}

\def\<{\langle}
\def\>{\rangle}
\def\lb{\label}

\def\ga{{\gamma}}
\def\Ga{{\Gamma}}

\def\sg{{\sigma}}

\def\diag{{\rm diag}}

\def\hb{\vrule height0.18cm width0.14cm $\,$}

\title{Time-Inconsistent Stochastic Linear-quadratic Differential Game}

\author{Qinglong Zhou$^{a} $\thanks{Partially supported by NSFC (No.11501330) and CPSF (Grant No. 2015M582071).
           E-mail: zhouqinglong@sdu.edu.cn}\quad
        Gaofeng Zong$^{a,b} $\thanks{Partially supported by NSFC (No. 11501325). E-mail: gf{\_}zong@126.com}, \\
\\$^{a}$ School of Mathematics,\\Shandong University, Jinan 250100, Shandong, China\\
  $^{b}$ School of Mathematics and Quantitative Economics,\\Shandong University of Finance and Economics, Jinan 250014, China\\
}

\begin{document}

\maketitle

\begin{abstract}
We consider a general time-inconsistent stochastic linear-quadratic differential game.
The time-inconsistency arises from the presence of quadratic terms of
the expected state as well as state-dependent term in the objective functionals.
We define an equilibrium strategy, which is different from the classical one,
and derived a sufficient conditions for equilibrium strategies
via a system of forward-backward stochastic differential equations.
When the state is one-dimensional and the coefficients are all deterministic,
we find an explicit equilibrium strategy.
The uniqueness of such equilibrium strategy is given.
\\

\\

\end{abstract}

\renewcommand{\theequation}{\thesection.\arabic{equation}}

\setcounter{equation}{0}
\setcounter{figure}{0}
\section{Introduction}
\label{sec:1}

Time inconsistency in dynamic decision making is often observe in social systems and daily life.
Motivated by practical applications, especially in mathematical economics and finance,
time-inconsistency control problems have recently attracted considerable research interest and efforts
attempting to seek equilibrium, instead of optimal, controls.
At a conceptual level, the idea is that a decision made by the controller at every instant of time is considered
as a game against all the decisions made by the future incarnations of the controller.
An ``equilibrium" control is therefore one such that any deviation from it at any time instant will be worse off.
The study on time inconsistency by economists can be dated back to
Stroz \cite{Stroz} and Phelps (\cite{Phelps,Phelps1}) in models with
discrete time (see \cite{Krusell} and \cite{Laibson} for further
developments), and adapted by Karp (\cite{Kar2,KaL1}), and by Ekeland
and Lazrak (\cite{Eke1,EkL,EkL1,EkL2,EKS,ELZ}) to the case of
continuous time.
In the LQ control problems, Yong \cite{Yon} studied a time-inconsistent deterministic model and derived equilibrium controls via some integral equations.

It is natural to study time inconsistency
in the stochastic models.
Ekeland and Pirvu \cite{EkP} studied the non-exponential discounting which leads to time inconsistency in an agent's
investment-consumption policies in a Merton model.
Grenadier and Wang \cite{GrW} also studied the hyperbolic discounting problem in an optimal stopping model.
In a Markovian systems, Bj\"{o}rk and Murgoci \cite{BjM} proposed a definition of a general stochastic
control problem with time inconsistent terms,
and proposed some sufficient condition for a control to be solution by a system of integro-differential equations.
They constructed some solutions for some examples including an LQ one,
but it looks very hard to find not-to-harsh condition on parameters to ensure the existence of a solution.
Bj\"{o}rk, Murgoci and Zhou \cite{BMZ} also constructed an equilibrium for a mean-variance portfolio selection
with state-dependent risk aversion.
Basak and Chabakauri \cite{BaC} studied the mean-variance portfolio selection problem and got more details on
the constructed solution.
Hu, Jin and Zhou \cite{HJZ,HJZ2} studied the general LQ
control problem with time inconsistent terms in a non-Markovian system
and constructed an unique equilibrium for quite general LQ control problem,
including a non-Markovian system.

To the best of our knowledge, most of the time-inconsistent problems are associated with
the control problems though we use the game formulation to define its equilibrium.
In the problems of game theory,
the literatures about time inconsistency is little \cite{Ben, Mar}.
However, the definitions of equilibrium strategies in the above two papers
are based on some corresponding control problems like before.
In this paper,
we formulate a general stochastic LQ differential game,
where the objective functional of each player include both a quadratic term of the expected state
and a state-dependent term.
These non-standard terms each introduces time inconsistency
into the problem in somewhat different ways.
We define our equilibrium via open-loop controls.
Then we derive
a general sufficient condition for equilibrium strategies through a system of forward-backward
stochastic differential equations (FBSDEs).
An intriguing feature of these FBSDEs is
that a time parameter is involved; so these form a flow of FBSDEs.
When the state
process is scalar valued and all the coefficients are deterministic functions of time,
we are able to reduce this flow of FBSDEs into several Riccati-like ODEs.
Comparing to the ODEs in \cite{HJZ}, though the state
process is scalar valued,
the unknowns are matrix-valued because of two players.
Therefore, such ODEs are harder to solve than those of \cite{HJZ}.
Under some more stronger conditions, we
obtain explicitly an equilibrium strategy, which turns out to be a linear feedback.
We also prove that the equilibrium strategy we obtained is unique.

The rest of the paper is organized as follows.
The next section is devoted to the formulation of our problem and the definition of equilibrium strategy.
In Section 3, we apply the spike variation technique to derive a flow of FBSEDs and a sufficient condition
of equilibrium strategies.
Based on this general results,
we solve in Section 4 the case when the state is one dimensional and all the coefficients are deterministic.
The uniqueness of such equilibrium strategy is also proved in this section.

\vspace{2mm}

\setcounter{equation}{0}
\section{Problem setting}
Let $T>0$ be  the end of a finite time horizon, and let
$(W_t)_{0\le t\le{T}}=(W_t^1,...,W_t^d)_{0\le t\le{T}}$ be a $d$-dimensional Brownian motion
on a probability space $(\Omega, \mathcal{F},\mathbb{P})$.
Denote by $(\mathcal{F}_t)$ the augmented filtration generated by $(W_t)$.

As in \cite{HJZ}, let $\mathbb{S}^n$ be the set of symmetric $n\times n$ real matrices;
$L_{\mathcal{F}}^2(\Omega,\mathbb{R}^l)$ be the set of square-integrable random variables;
$L_{\mathcal{F}}^2(t,T;\mathbb{R}^n)$ be the set of $\{\mathcal{F}_s\}_{s\in[t,T]}$-adapted square-integrable processes;
and $L_{\mathcal{F}}^2(\Omega;C(t,T;\mathbb{R}^n))$ be the set of continuous $\{\mathcal{F}_s\}_{s\in[t,T]}$-adapted
square-integrable processes.

We consider a continuous-time, $n$-dimensional nonhomogeneous linear controlled system (cf. \cite{HJZ})
\be
dX_s=[A_sX_s+B_{1,s}'u_{1,s}+B_{2,s}'u_{2,s}+b_s]ds+\sum_{j=1}^d[C_s^jX_s+D_{1,s}^ju_{1,s}+D_{2,s}^ju_{2,s}+\sigma_s^j]dW_s^j, \quad X_0=x_0.
\ee
Here $A$ is a bounded deterministic function on $[0,T]$ with value in $\mathbb{R}^{n\times n}$.
The other parameters $B_1, B_2, C, D_1, D_2$ are all essentially bounded adapted processes on $[0,T]$
with values in $\mathbb{R}^{l\times n},\mathbb{R}^{l\times n},\mathbb{R}^{n\times n},\mathbb{R}^{n\times l},\mathbb{R}^{n\times l}$, respectively; $b$ and $\sigma^j$ are stochastic processes in $L^2_{\mathcal{F}}(0,T;\mathbb{R}^n)$.
The processes $u_i\in L^2_{\mathcal{F}}(0,T;\mathbb{R}^l),\ i=1,2$ are the controls,
and $X$ is the state process valued in $\mathbb{R}^{n}$.
Finally, $x_0\in\mathbb{R}^{n}$ is the initial state.
It is obvious that for any controls $u_i\in L^2_{\mathcal{F}}(0,T;\mathbb{R}^l),\ i=1,2$,
there exists a unique solution $X\in L^2_{\mathcal{F}}(\Omega, C(0,T;\mathbb{R}^n))$.

As time evolves, we need to consider the controlled system starting from time $t\in[0,T]$ and
state $x_t\in L^2_{\mathcal{F}_t}(\Omega;\mathbb{R}^n)$:
\be
dX_s=[A_sX_s+B_{1,s}'u_{1,s}+B_{2,s}'u_{2,s}+b_s]ds+\sum_{j=1}^d[C_s^jX_s+D_{1,s}^ju_{1,s}+D_{2,s}^ju_{2,s}+\sigma_s^j]dW_s^j, \quad X_t=x_t.
\ee
For any controls $u_i\in L^2_{\mathcal{F}}(0,T;\mathbb{R}^l),\ i=1,2$, there exists a unique solution
$X^{t,x_t,u_1,u_2}\in L^2_{\mathcal{F}}(\Omega, C(0,T;\mathbb{R}^n))$.

We consider a two-person differential game problem.
At any time $t$ with the system state $X_t=x_t$, the $i$-th ($i=1,2$) person's aim is to minimize her cost (if maximize, we can times the following function by $-1$):
\begin{eqnarray}
J_i(t,x_t;u_1,u_2)&=&\frac{1}{2}\mathbb{E}_t\int_t^T[\<Q_{i,s}X_s,X_s\>+\<R_{i,s}u_{i,s},u_{i,s}\>]ds
+\frac{1}{2}\mathbb{E}_t[\<G_iX_T,X_T\>]
\nonumber\\
&&-\frac{1}{2}\<h_i\mathbb{E}_t[X_T],\mathbb{E}_t[X_T]\>
-\<\lambda_ix_t+\mu_i,\mathbb{E}_t[X_T]\>
\end{eqnarray}
over $u_1,u_2\in L_{\mathcal{F}}^2(t,T;\mathbb{R}^l)$, where $X=X^{t,x_t,u_1,u_2}$, and $\mathbb{E}_t[\cdot]=\mathbb{E}[\cdot|\mathcal{F}_t]$.
Here, for $i=1,2$, $Q_i$ and $R_i$ are both given essentially bounded adapted process on $[0,T]$ with values in $\mathbb{S}^n$
and $\mathbb{S}^l$, respectively, $G_i,h_i,\lambda_i,\mu_i$ are all constants in $\mathbb{S}^n$, $\mathbb{S}^n$,
$\mathbb{R}^{n\times n}$ and $\mathbb{R}^n$, respectively.
Furthermore, we assume that $Q_i, R_i$ are non-negative definite almost surely and $G_i$ are non-negative definite.

Given a control pair $(u_1^*,u_2^*)$. For any $t\in[0,T),\epsilon>0$, and $v_1,v_2\in L_{\mathcal{F}_t}^2(\Omega,\mathbb{R}^l)$, define
\begin{equation}\label{spiked.control}
u_{i,s}^{t,\epsilon,v_i}=u_{i,s}^*+v_i{\bf 1}_{s\in[t,t+\epsilon)},\quad\quad s\in[t,T],\ i=1,2.
\end{equation}

\begin{definition}
Let $(u_1^*,u_2^*)\in L_{\mathcal{F}}^2(0,T;\mathbb{R}^l)\times L_{\mathcal{F}}^2(0,T;\mathbb{R}^l)$ be a given strategy pair, and let $X^*$ be the state process corresponding to $(u_1^*,u_2^*)$. The strategy pair $(u_1^*,u_2^*)$
is called an equilibrium if
\begin{eqnarray}
\lim_{\epsilon\downarrow0}\frac{J_1(t,X_t^*;u_1^{t,\epsilon,v_1},u_2^*)-J_1(t,X_t^*;u_1^*,u_2^*)}{\epsilon}\ge0,\label{cond1}
\\
\lim_{\epsilon\downarrow0}\frac{J_2(t,X_t^*;u_1^*,u_2^{t,\epsilon,v_2})-J_2(t,X_t^*;u_1^*,u_2^*)}{\epsilon}\ge0,\label{cond2}
\end{eqnarray}
\end{definition}
where $u_i^{t,\epsilon,v_i}, i=1,2$ are defined by (\ref{spiked.control}), for any $t\in[0,T)$
and $v_1,v_2\in L_{\mathcal{F}_t}^2(\Omega,\mathbb{R}^l)$.

{\bf Remark. }The ``$\ge$'' in (\ref{cond1})-(\ref{cond2}) because of each person want to minimize his/her cost as we claimed before.
The above definition means that, in each time $t$, the equilibrium is a static Nash equilibrium
in a corresponding game.

\section{Sufficient conditions}

Let $(u_1^*,u_2^*)$ be a fixed strategy pair, and let $X^*$ be the corresponding state process.
For any $t\in[0,T)$, define in the time interval $[t,T]$ the processes
$(p_i(\cdot;t), (k_i^j(\cdot;t)_{j=1,2,...,d}))\in L_{\mathcal{F}}^2(t,T;\mathbb{R}^n)\times(L_{\mathcal{F}}^2(t,T;\mathbb{R}^n))^d$ and
$(P_i(\cdot;t), (K_i^j(\cdot;t)_{j=1,2,...,d}))\in L_{\mathcal{F}}^2(t,T;\mathbb{S}^n)\times(L_{\mathcal{F}}^2(t,T;\mathbb{S}^n))^d$ for $i=1,2$
are the solutions to the following equations:
\begin{eqnarray}
&&\left\{
\begin{array}{l}
    dp_i(s;t)=-[A_s'p_i(s;t)+\sum_{j=1}^d(C_s^j)'k_i^j(s;t)+Q_{i,s}X_s^*]ds+\sum_{j=1}^dk_i^j(s;t)dW_s^j,\quad s\in[t,T],
    \\
    p_i(T;t)=G_iX_T^*-h_i\mathbb{E}_t[X_T^*]-\lambda_iX_t^*-\mu_i,
    \end{array}
\right.\label{BSDE.pk}
\\
&&\left\{
\begin{array}{l}
    dP_i(s;t)=-\bigg\{A_s'P_i(s;t)+P_i(s;t)A_s+Q_{i,s}+\sum_{j=1}^d[(C_s^j)'P_i(s;t)C_s^j+(C_s^j)'K_i^j(s;t)+K_i^j(s;t)C_s^j]\bigg\}ds
    \\
    \quad\quad\quad\quad+\sum_{j=1}^dK_i^j(s;t)dW_i^j,\quad s\in[t,T],
    \\
    P_i(T;t)=G_i,
    \end{array}
\right.\label{BSDE.PK}
\end{eqnarray}
for $i=1,2$.
From the assumption that $Q_i$ and $G_i$ are non-negative definite, it follows that $P_i(s;t)$ are non-negative definite
for $i=1,2$.

\begin{proposition}\label{Prop:variations}
For any $t\in[0,T),\epsilon>0$, and $v_1,v_2\in L_{\mathcal{F}_t}^2(\Omega,\mathbb{R}^l)$, define
$u_i^{t,\epsilon,v_i}, i=1,2$ by (\ref{spiked.control}). Then
\begin{eqnarray}
J_1(t,X_t^*;u_1^{t,\epsilon,v_1},u_2^*)-J_1(t,X_t^*;u_1^*,u_2^*)=\mathbb{E}_t\int_t^{t+\epsilon}\bigg\{\<\Lambda_1(s;t),v_1\>+\frac{1}{2}\<H_1(s;t)v_1,v_1\>
\bigg\}ds+o(\epsilon),\label{DJ1}
\\
J_2(t,X_t^*;u_1^*,u_2^{t,\epsilon,v_2})-J_2(t,X_t^*;u_1^*,u_2^*)=\mathbb{E}_t\int_t^{t+\epsilon}\bigg\{\<\Lambda_2(s;t),v_2\>+\frac{1}{2}\<H_2(s;t)v_2,v_2\>
\bigg\}ds+o(\epsilon),\label{DJ2}
\end{eqnarray}
where $\Lambda_i(s;t)=B_{i,s}p_i(s;t)+\sum_{j=1}^d(D_{i,s}^j)'k_i^j(s;t)+R_{i,s}u_{i,s}^*$ and
$H_i(s;t)=R_{i,s}+\sum_{j=1}^d(D_{i,s}^j)'P_i(s;t)D_{i,s}^j$ for $i=1,2$.
\end{proposition}

{\bf Proof. }Let $X^{t,\epsilon,v_1,v_2}$ be the state process corresponding to $u_i^{t,\epsilon,v_i},i=1,2$.
Then by standard perturbation approach (cf. \cite{Peng,HJZ} or pp. 126-128 of \cite{YoZ}), we have
\begin{equation}
X_s^{t,\epsilon,v_1,v_2}=X_s^*+Y_s^{t,\epsilon,v_1,v_2}+Z_s^{t,\epsilon,v_1,v_2},\quad s\in[t,T],
\end{equation}
where $Y\equiv Y^{t,\epsilon,v_1,v_2}$ and $Z\equiv Z^{t,\epsilon,v_1,v_2}$ satisfy
\begin{eqnarray}
\left\{
\begin{array}{l}
    dY_s=A_sY_sds+\sum_{j=1}^d[C_s^jY_s+D_{1,s}^jv_1{\bf 1}_{s\in[t,t+\epsilon)}+D_{2,s}^jv_2{\bf 1}_{s\in[t,t+\epsilon)}]dW_s^j,\quad s\in[t,T],
    \\
    Y_t=0,
    \end{array}
\right.
\\
\left\{
\begin{array}{l}
    dZ_s=[A_sZ_s+B_{1,s}'v_1{\bf 1}_{s\in[t,t+\epsilon)}+B_{2,s}'v_2{\bf 1}_{s\in[t,t+\epsilon)}]ds
         +\sum_{j=1}^dC_s^jZ_sdW_s^j,\quad s\in[t,T],
    \\
    Z_t=0.
    \end{array}
\right.
\end{eqnarray}
Moreover, by Theorem 4.4 in \cite{YoZ}, we have
\begin{equation}
\mathbb{E}_t\bigg[\sup_{s\in[t,T)}|Y_s|^2\bigg]=O(\epsilon),\quad
\mathbb{E}_t\bigg[\sup_{s\in[t,T)}|Z_s|^2\bigg]=O(\epsilon^2).
\end{equation}

With $A$ being deterministic, it follows from the dynamics of $Y$ that, for any $s\in[t,T]$, we have
\begin{equation}
\mathbb{E}_t[Y_s]=\int_t^s\mathbb{E}_t[A_sY_\tau]d\tau=\int_t^sA_s\mathbb{E}_t[Y_\tau]d\tau.
\end{equation}
Hence we conclude that
\begin{equation}
\mathbb{E}_t[Y_s]=0\quad s\in[t,T].
\end{equation}

By these estimates, we can calculate
\begin{eqnarray}\label{payoff.variation}
&&J_i(t,X_t^*;u_1^{t,\epsilon,v_1},u_2^{t,\epsilon,v_2})-J_i(t,X_t^*;u_1^*,u_2^*)
\nonumber\\
&&=\frac{1}{2}\mathbb{E}_t\int_t^T[\<Q_{i,s}(2X_s^*+Y_s+Z_s),Y_s+Z_s\>+\<R_{i,s}(2u_i^*+v_i),v_i\>{\bf 1}_{s\in[t,t+\epsilon)}]ds
\nonumber\\
&&\quad+\mathbb{E}_t[\<G_iX_T^*,Y_T+Z_T\>]+\frac{1}{2}\mathbb{E}_t[\<G_i(Y_T+Z_T),Y_T+Z_T\>]
\nonumber\\
&&\quad-\<h_i\mathbb{E}_t[X_T^*]+\lambda_iX_t^*+\mu_i,\mathbb{E}_t[Y_T+Z_T]\>-\frac{1}{2}\<h_i\mathbb{E}_t[Y_T+Z_T],\mathbb{E}_t[Y_T+Z_T]\>
\nonumber\\
&&=\frac{1}{2}\mathbb{E}_t\int_t^T[\<Q_{i,s}(2X_s^*+Y_s+Z_s),Y_s+Z_s\>+\<R_{i,s}(2u_i^*+v_i),v_i\>{\bf 1}_{s\in[t,t+\epsilon)}]ds
\nonumber\\
&&\quad+\mathbb{E}_t[\<G_iX_T^*-h_i\mathbb{E}_t[X_T^*]-\lambda_iX_t^*-\mu_i,Y_T+Z_T\>+\frac{1}{2}\<G_i(Y_T+Z_T),Y_T+Z_T\>]+o(\epsilon).
\end{eqnarray}
Recalling that $(p_i(\cdot;t),k_i(\cdot;t))$ and $(P_i(\cdot;t),K_i(\cdot;t))$ solve, respectively, (\ref{BSDE.pk}) and
(\ref{BSDE.PK}) for $i=1,2$, we have
\begin{eqnarray}
&&\mathbb{E}_t[\<G_iX_T^*-h_i\mathbb{E}_t[X_T^*]-\lambda_iX_t^*-\mu_i,Y_T+Z_T\>]
\nonumber\\
&&=\mathbb{E}_t[\<p_i(T;t),Y_T+Z_T\>]
\nonumber\\
&&=\mathbb{E}_t\bigg[\int_t^Td\<p_i(s;t),Y_s+Z_s\>\bigg]
\nonumber\\
&&=\mathbb{E}_t\int_t^T\bigg[\<p_i(s;t),A_s(Y_s+Z_s)+B_{1,s}'v_1{\bf 1}_{s\in[t,t+\epsilon)}+B_{2,s}'v_2{\bf 1}_{s\in[t,t+\epsilon)}\>
\nonumber\\
&&\quad-\<A_s'p_i(s;t)+\sum_{j=1}^d(C_s^j)'k_i^j(s;t)+Q_{i,s}X_s^*,Y_s+Z_s\>
\nonumber\\
&&\quad+\sum_{j=1}^d\<k_i^j(s;t),C_s^j(Y_s+Z_s)+D_{1,s}^jv_1{\bf 1}_{s\in[t,t+\epsilon)}+D_{2,s}^jv_2{\bf 1}_{s\in[t,t+\epsilon)}\>\bigg]ds
\nonumber\\
&&=\mathbb{E}_t\int_t^T\bigg[\<-Q_{i,s}X_s^*\>+\bigg\<B_{1,s}p_i(s;t)+\sum_{j=1}^d(D_{1,s}^j)'k_i^j(s;t),v_1{\bf 1}_{s\in[t,t+\epsilon)}\bigg\>
\nonumber\\
&&\quad+\bigg\<B_{2,s}p_i(s;t)+\sum_{j=1}^d(D_{2,s}^j)'k_i^j(s;t),v_2{\bf 1}_{s\in[t,t+\epsilon)}\bigg\>
\bigg]ds
\end{eqnarray}
and
\begin{eqnarray}\label{term.2}
&&\mathbb{E}_t[\frac{1}{2}\<G_i(Y_T+Z_T),Y_T+Z_T\>]
\nonumber\\
&&=\mathbb{E}_t[\frac{1}{2}\<P_i(T;t)(Y_T+Z_T),Y_T+Z_T\>]
\nonumber\\
&&=\mathbb{E}_t\bigg[\int_t^Td\<P_i(s;t)(Y_s+Z_s),Y_s+Z_s\>\bigg]
\nonumber\\
&&=\mathbb{E}_t\int_t^T\bigg\{\<P_i(s;t)(Y_s+Z_s),A_s(Y_s+Z_s)+B_{1,s}'v_1{\bf 1}_{s\in[t,t+\epsilon)}+B_{2,s}'v_2{\bf 1}_{s\in[t,t+\epsilon)}\>
\nonumber\\
&&\quad+\<P_i(s;t)[A_s(Y_s+Z_s)+B_{1,s}'v_1{\bf 1}_{s\in[t,t+\epsilon)}+B_{2,s}'v_2{\bf 1}_{s\in[t,t+\epsilon)}],Y_s+Z_s\>
\nonumber\\
&&\quad-\<[A_s'P_i(s;t)+P_i(s;t)A_s+Q_{i,s}+\sum_{j=1}^d((C_s^j)'P_i(s;t)C_s^j+(C_s^j)'K_i^j(s;t)+K_i^j(s;t)C_s^j)](Y_s+Z_s),Y_s+Z_s\>
\nonumber\\
&&\quad+\sum_{j=1}^d\<K_i^j(s;t)(Y_s+Z_s),C_s^j(Y_s+Z_s)+D_{1,s}^jv_1{\bf 1}_{s\in[t,t+\epsilon)}+D_{2,s}^jv_2{\bf 1}_{s\in[t,t+\epsilon)}\>
\nonumber\\
&&\quad+\sum_{j=1}^d\<K_i^j(s;t)[C_s^j(Y_s+Z_s)+D_{1,s}^jv_1{\bf 1}_{s\in[t,t+\epsilon)}+D_{2,s}^jv_2{\bf 1}_{s\in[t,t+\epsilon)}],Y_s+Z_s\>
\nonumber\\
&&\quad+\sum_{j=1}^d\<P_i(s;t)[C_s^j(Y_s+Z_s)+D_{1,s}^jv_1{\bf 1}_{s\in[t,t+\epsilon)}+D_{2,s}^jv_2{\bf 1}_{s\in[t,t+\epsilon)}],\nonumber\\
&&\qquad\qquad C_s^j(Y_s+Z_s)+D_{1,s}^jv_1{\bf 1}_{s\in[t,t+\epsilon)}+D_{2,s}^jv_2{\bf 1}_{s\in[t,t+\epsilon)}\>
\bigg\}ds
\nonumber\\
&&=\mathbb{E}_t\int_t^T\bigg[-\<Q_{i,s}(Y_s+Z_s),Y_s+Z_s\>
\nonumber\\
&&\quad+\sum_{j=1}^d\<P_i(s;t)[D_{1,s}^jv_1+D_{2,s}^jv_2],D_{1,s}^jv_1+D_{2,s}^jv_2\>{\bf 1}_{s\in[t,t+\epsilon)}
\bigg]ds+o(\epsilon)
\end{eqnarray}

Combining (\ref{payoff.variation})-(\ref{term.2}), we have
\begin{eqnarray}\label{payoff.variation.2}
&&J_i(t,X_t^*;u_1^{t,\epsilon,v_1},u_2^{t,\epsilon,v_2})-J_i(t,X_t^*;u_1^*,u_2^*)
\nonumber\\
&&=\mathbb{E}_t\int_t^T\bigg[\frac{1}{2}\<R_{i,s}(2u_i^*+v_i),v_i\>{\bf 1}_{s\in[t,t+\epsilon)}
+\bigg\<B_{1,s}p_i(s;t)+\sum_{j=1}^d(D_{1,s}^j)'k_i^j(s;t),v_1{\bf 1}_{s\in[t,t+\epsilon)}\bigg\>
\nonumber\\
&&\quad+\bigg\<B_{2,s}p_i(s;t)+\sum_{j=1}^d(D_{2,s}^j)'k_i^j(s;t),v_2{\bf 1}_{s\in[t,t+\epsilon)}\bigg\>
\nonumber\\
&&\quad+\frac{1}{2}\sum_{j=1}^d\<P_i(s;t)[D_{1,s}^jv_1+D_{2,s}^jv_2],D_{1,s}^jv_1+D_{2,s}^jv_2\>{\bf 1}_{s\in[t,t+\epsilon)}
\bigg]ds+o(\epsilon).
\end{eqnarray}
Take $i=1$, we let $v_2=0$, then $u_2^{t,\epsilon,v_2}=u_2^*$, from (\ref{payoff.variation.2}), we obtain
\begin{eqnarray}
&&J_1(t,X_t^*;u_1^{t,\epsilon,v_1},u_2^*)-J_1(t,X_t^*;u_1^*,u_2^*)
\nonumber\\
&&=\mathbb{E}_t\int_t^T\bigg\{\bigg\<R_{1,s}u_1^*+B_{1,s}p_1(s;t)+\sum_{j=1}^d(D_{1,s}^j)'k_1^j(s;t),v_1{\bf 1}_{s\in[t,t+\epsilon)}\bigg\>
\nonumber\\
&&\quad+\frac{1}{2}\bigg\<\bigg[R_{1,s}+\sum_{j=1}^d(D_{1,s}^j)'P_1(s;t)D_{1,s}^j\bigg]v_1,v_1\bigg\>
\bigg\}ds
\nonumber\\
&&=\mathbb{E}_t\int_t^{t+\epsilon}\bigg\{\<\Lambda_1(s;t),v_1\>+\frac{1}{2}\<H_1(s;t)v_1,v_1\>
\bigg\}ds+o(\epsilon).
\end{eqnarray}
This prove (\ref{DJ1}), and similarly, we obtain (\ref{DJ2}).
\hb

Because of $R_{i,s}$ and $P_i(s;t),i=1,2$ are non-negative definite, $H_i(s;t),\ i=1,2$ are also non-negative definite.
In view of (\ref{DJ1})-(\ref{DJ2}), a sufficient condition for an equilibrium is
\begin{equation}\label{sufficient.condition}
\mathbb{E}_t\int_t^T|\Lambda_i(s;t)|ds<+\infty,
\quad
\lim_{s\downarrow t}\mathbb{E}_t[\Lambda_i(s;t)]=0\ a.s.\ \forall t\in[0,T],
\quad i=1,2.
\end{equation}

Similar to Proposition 3.3 of \cite{HJZ2},
we have the following lemma:
\begin{lemma}\label{Lm:kp.simplify}
For any triple of state and control processes $(X^*,u_1^*,u_2^*)$,
the solution to (\ref{BSDE.pk}) in $L^2(0,T;\mathbb{R}^n)\times (L^2(0,T;\mathbb{R}^n))^d$ satisfies
$k_i(s;t_1)=k_i(s;t_2)$ for a.e. $s\ge\max\{t_1,t_2\},\;i=1,2$.
Furthermore, there exist $\rho_i\in L^2(0,T;\mathbb{R}^l)$,$\delta_i\in L^2(0,T;\mathbb{R}^{l\times n})$
and $\xi_i\in L^2(\Omega;C(0,T;\mathbb{R}^n))$, such that
\be
\Lambda_i(s;t)=\rho_i(s)+\delta_i(s)\xi_i(t),\qquad i=1,2.
\ee
\end{lemma}
Therefore, we have another characterization for equilibrium strategies:
\begin{theorem}\label{Th:equilibrium.strategy.with.new.condition}
Given a strategy pair $(u_1^*,u_2^*)\in L^2(0,T;\mathbb{R}^l)\times L^2(0,T;\mathbb{R}^l)$.
Denote $X^*$ as the state process,
and $(p_i(\cdot;t), (k_i^j(\cdot;t)_{j=1,2,...,d}))\in L_{\mathcal{F}}^2(t,T;\mathbb{R}^n)\times(L_{\mathcal{F}}^2(t,T;\mathbb{R}^n))^d$ as the unique solution
for the BSDE (\ref{BSDE.pk}), with $k_i(s)=k_i(s;t)$ according to Lemma \ref{Lm:kp.simplify}
for $i=1,2$ respectively.
For $i=1,2$, letting
\be
\Lambda_i(s,t)=B_{i,s}p_i(s;t)+\sum_{j=1}^d(D_{j,s})'k(s;t)^j+R_{i,s}u_{i,s}^*,\quad s\in[t,T],
\ee
then $u^*$ is an equilibrium strategy if and only if
\be\label{new.condition}
\Lambda_i(t,t)=0,\;a.s.,\; a.e.\;t\in[0,T],\quad i=1,2.
\ee
\end{theorem}

{\bf Proof.} The proof is by Lemma 3.4 of \cite{HJZ2} and Theorem \ref{Th:equilibrium.strategy}.\hb

The following is the main general result for the time-inconsistent stochastic LQ differential game.

\begin{theorem}\label{Th:equilibrium.strategy}
A strategy pair $(u_1^*,u_2^*)\in L_{\mathcal{F}}^2(0,T;\mathbb{R}^l)\times L_{\mathcal{F}}^2(0,T;\mathbb{R}^l)$ is an equilibrium strategy pair if the following two conditions hold for any time $t$:

(i) The system of SDEs
\begin{eqnarray}\label{FBSDEs}
\left\{
\begin{array}{l}
    dX_s^*=[A_sX_s^*+B_{1,s}'u_{1,s}^*+B_{2,s}'u_{2,s}^*+b_s]ds+\sum_{j=1}^d[C_s^jX_s^*+D_{1,s}^ju_{1,s}^*+D_{2,s}^ju_{2,s}^*+\sigma_s^j]dW_s^j,
    \\
    X_0^*=x_0,
    \\
    dp_1(s;t)=-[A_s'p_1(s;t)+\sum_{j=1}^d(C_s^j)'k_1^j(s;t)+Q_{1,s}X_s^*]ds+\sum_{j=1}^dk_1^j(s;t)dW_s^j,\quad s\in[t,T],
    \\
    p_1(T;t)=G_1X_T^*-h_1\mathbb{E}_t[X_T^*]-\lambda_1X_t^*-\mu_1,
    \\
    dp_2(s;t)=-[A_s'p_2(s;t)+\sum_{j=1}^d(C_s^j)'k_2^j(s;t)+Q_{2,s}X_s^*]ds+\sum_{j=1}^dk_2^j(s;t)dW_s^j,\quad s\in[t,T],
    \\
    p_2(T;t)=G_2X_T^*-h_2\mathbb{E}_t[X_T^*]-\lambda_2X_t^*-\mu_2,
    \end{array}
\right.
\end{eqnarray}
admits a solution $(X^*,p_1,k_1,p_2,k_2)$;

(ii) $\Lambda_i(s;t)=R_{i,s}u_{i,s}^*+B_{i,s}p_i(s;t)+\sum_{j=1}^d(D_{i,s}^j)'k_i^j(s;t),i=1,2$ satisfy condition (\ref{new.condition}).
\end{theorem}

{\bf Proof. }Given a strategy pair $(u_1^*,u_2^*)\in L_{\mathcal{F}}^2(0,T;\mathbb{R}^l)\times L_{\mathcal{F}}^2(0,T;\mathbb{R}^l)$ satisfying (i) and (ii), then for any
$v_1,v_2\in L_{\mathcal{F}_t}^2(\Omega,\mathbb{R}^l)$,
define $\Lambda_i,H_i, i=1,2$ as in Proposition \ref{Prop:variations}.
We have
\begin{eqnarray}
&&\lim_{\epsilon\downarrow0}\frac{J_1(t,X_t^*;u_1^{t,\epsilon,v_1},u_2^*)-J_1(t,X_t^*;u_1^*,u_2^*)}{\epsilon}
\nonumber\\
&&=\lim_{\epsilon\downarrow0}\frac{\mathbb{E}_t\int_t^{t+\epsilon}\bigg\{\<\Lambda_1(s;t),v_1\>+\frac{1}{2}\<H_1(s;t)v_1,v_1\>
\bigg\}ds}{\epsilon}
\nonumber\\
&&\ge\lim_{\epsilon\downarrow0}\frac{\mathbb{E}_t\int_t^{t+\epsilon}\<\Lambda_1(s;t),v_1\>ds}{\epsilon}
\nonumber\\
&&=0,
\end{eqnarray}
proving the first condition of Definition 2.1, and the proof of the second condition is similar.
\hb

Theorem \ref{Th:equilibrium.strategy}
involve the existence of solutions to a flow of FBSDEs
along with other conditions.
The system (\ref{FBSDEs}) is more complicated than system (3.6) in \cite{HJZ}.
As declared in \cite{HJZ}, ``proving the general existence for this type of FBSEs remains an outstanding open problem",
it is also true for our system (\ref{FBSDEs}).

In the rest of this paper, we will focus on the case when $n=1$.
When $n=1$, the state process $X$ is a scalar-valued rocess evolving by the dynamics
\be
dX_s=[A_sX_s+B_{1,s}'u_{1,s}+B_{2,s}'u_{2,s}+b_s]ds+[C_sX_s+D_{1,s}u_{1,s}+D_{2,s}u_{2,s}+\sigma_s]'dW_s, \quad X_0=x_0,
\ee
where $A$ is a bounded deterministic scalar function on $[0,T]$.
The other parameters $B,C,D$ are all essentially bounded and $\mathcal{F}_t$-adapted processes on $[0,T]$
with values in $\mathbb{R}^l,\mathbb{R}^d,\mathbb{R}^{d\times l}$, respectively.
Moreover, $b\in L_{\mathcal{F}}^2(0,T;\mathbb{R})$ and $\sg\in L_{\mathcal{F}}^2(0,T;\mathbb{R}^d)$.

In this case, the adjoint equations for the equilibrium strategy become
\begin{eqnarray}
&&\left\{
\begin{array}{l}
    dp_i(s;t)=-[A_s'p_i(s;t)+(C_s)'k_i(s;t)+Q_{i,s}X_s^*]ds+k_i(s;t)'dW_s,\quad s\in[t,T],
    \\
    p_i(T;t)=G_iX_T^*-h_i\mathbb{E}_t[X_T^*]-\lambda_iX_t^*-\mu_i,
    \end{array}
\right.\label{BSDE.pk.n=1}
\\
&&\left\{
\begin{array}{l}
    dP_i(s;t)=-[(2A_s+|C_s|^2)P_i(s;t)+2C_s'K(s;t)+Q_{i,s}]ds
+K_i(s;t)'dW_s,\quad s\in[t,T],
    \\
    P_i(T;t)=G_i,
    \end{array}
\right.\label{BSDE.PK.n=1}
\end{eqnarray}
for $i=1,2$.
For convenience, we also state here the $n=1$ version of Theorem \ref{Th:equilibrium.strategy}:
\begin{theorem}
A strategy pair $(u_1^*,u_2^*)\in L_{\mathcal{F}}^2(0,T;\mathbb{R}^l)\times L_{\mathcal{F}}^2(0,T;\mathbb{R}^l)$ is an equilibrium strategy pair if,
for any time $t\in[0,T)$,

(i) The system of SDEs
\begin{eqnarray}\label{FBSDEs.n=1}
\left\{
\begin{array}{l}
    dX_s^*=[A_sX_s^*+B_{1,s}'u_{1,s}^*+B_{2,s}'u_{2,s}^*+b_s]ds+[C_sX_s^*+D_{1,s}u_{1,s}^*+D_{2,s}u_{2,s}^*+\sigma_s]'dW_s,
    \\
    X_0^*=x_0,
    \\
    dp_1(s;t)=-[A_sp_1(s;t)+(C_s)'k_1(s;t)+Q_{1,s}X_s^*]ds+k_1(s;t)'dW_s,\quad s\in[t,T],
    \\
    p_1(T;t)=G_1X_T^*-h_1\mathbb{E}_t[X_T^*]-\lambda_1X_t^*-\mu_1,
    \\
    dp_2(s;t)=-[A_sp_2(s;t)+(C_s)'k_2(s;t)+Q_{2,s}X_s^*]ds+k_2(s;t)'dW_s,\quad s\in[t,T],
    \\
    p_2(T;t)=G_2X_T^*-h_2\mathbb{E}_t[X_T^*]-\lambda_2X_t^*-\mu_2,
    \end{array}
\right.
\end{eqnarray}
admits a solution $(X^*,p_1,k_1,p_2,k_2)$;

(ii) $\Lambda_i(s;t)=R_{i,s}u_{i,s}^*+B_{i,s}p_i(s;t)+(D_{i,s})'k_i(s;t),i=1,2$ satisfy condition (\ref{new.condition}).
\end{theorem}

\section{Existence and uniqueness of equilibrium strategy when coefficients are deterministic}

The unique solvability of (\ref{FBSDEs.n=1}) remains a challenging open problem even for the case $n=1$.
However, we are able to solve this problem when the parameters $A,B_1,B_2,C,D_1,D_2,b,\sg,Q_1,Q_2,R_1$
and $R_2$ are all deterministic functions.

Throughout this section we assume all the parameters are deterministic functions of $t$.
In this case, since $G_1,G_2$ has been also assumed to be deterministic,
the BSDEs (\ref{BSDE.PK.n=1}) turns out to be ODEs with solutions $K_i\equiv0$ and
$P_i(s;t)=G_ie^{\int_s^T(2A_u+|C_u|^2)du}+\int_s^Te^{\int_s^T(2A_u+|C_u|^2)du}Q_{i,v}dv$ for $i=1,2$.

\subsection{An intuitional idea and the uniqueness of the equilibrium strategy}

As in classical LQ control, we attempt to look for a linear feedback equilibrium strategy pair.
For such purpose, motivated by \cite{HJZ},
given any $t\in[0,T]$, we consider the following process:
\be\label{Ansatz}
p_i(s;t)=M_{i,s}X_s^*-N_{i,s}\mathbb{E}_t[X_s^*]-\Ga_{i,s}X_t^*+\Phi_{i,s},\;0\le t\le s\le T,\;\;i=1,2,
\ee
where $M_i,N_i,\Ga_i,\Phi_i$ are deterministic differentiable functions with $\dot{M}_i=m_i,\dot{N}_i=n_i,\dot{\Ga}_i=\ga_i$
and $\dot\Phi_i=\phi_i$ for $i=1,2$.
The advantage of this process is to separate the variables $X_s^*,\mathbb{E}_t[X_s^*]$ and $X_t^*$ in the
solutions $p_i(s;t),i=1,2$,
thereby reducing the complicated FBSDEs to some ODEs.

For any fixed $t$, applying Ito's formula to (\ref{Ansatz}) in the time variable $s$, we obtain, for $i=1,2$,
\bea\label{dpi.Ansatz}
dp_i(s;t)&=&\{M_{i,s}(A_sX_s^*+B_{1,s}'u_{1,s}^*+B_{2,s}'u_{2,s}^*+b_s)+m_{i,s}X_s^*-N_{i,s}\mathbb{E}_t[A_sX_s^*+B_{1,s}'u_{1,s}^*+B_{2,s}'u_{2,s}^*+b_s]
\nonumber\\
&&\qquad -n_{i,s}\mathbb{E}_t[X_s^*]-\ga_{i,s}X_t^*+\phi_{i,s}\}ds+M_{i,s}(C_sX_s^*+D_{1,s}u_{1,s}^*+D_{2,s}u_{2,s}^*+\sg_s)'dW_s.
\eea
Comparing the $dW_s$ term of $dp_i(s;t)$ in (\ref{FBSDEs.n=1}) and (\ref{dpi.Ansatz}), we have
\be\label{ki.Ansatz}
k_i(s;t)=M_{i,s}[C_sX_s^*+D_{1,s}u_{1,s}^*+D_{2,s}u_{2,s}^*+\sg_s],\;s\in[t,T],\quad i=1,2.
\ee
Notice that $k(s;t)$ turns out to be independent of $t$.

Putting the above expressions (\ref{Ansatz}) and (\ref{ki.Ansatz}) of $p_i(s;t)$ and $k_i(s;t),i=1,2$
into (\ref{new.condition}), we have
\be\label{equilibrium.equality.2}
R_{i,s}u_{i,s}^*+B_{i,s}[(M_{i,s}-N_{i,s}-\Ga_{i,s})X_s^*+\Phi_{i,s}]+D_{i,s}'M_{i,s}[C_sX_s^*+D_{1,s}u_{1,s}^*+D_{2,s}u_{2,s}^*+\sg_s]=0,
\;s\in[0,T],
\ee
for $i=1,2$.
Then we can formally deduce
\be\label{u.of.alpha.beta}
u_{i,s}^*=\alpha_{i,s}X_s^*+\beta_{i,s},\quad i=1,2.
\ee
Let $M_s=\diag(M_{1,s}I_l,M_{2,s}I_l),N_s=\diag(N_{1,s}I_l,N_{2,s}I_l),\Ga_s=\diag(\Ga_{1,s}I_l,\Ga_{2,s}I_l),\Phi_s=\diag(\Phi_{1,s}I_l,\Phi_{2,s}I_l)$,
$R_s=\diag(R_{1,s},R_{2,s}),B_s=\left(\matrix{B_{1,s}\cr B_{2,s}}\right), D_s=\left(\matrix{D_{1,s},\;D_{2,s}}\right)$,
$u_s^*=\left(\matrix{u_{1,s}^*\cr u_{2,s}^*}\right),\alpha_s=\left(\matrix{\alpha_{1,s}\cr \alpha_{2,s}}\right)$
and $\beta_s=\left(\matrix{\beta_{1,s}\cr \beta_{2,s}}\right)$.
Then from (\ref{equilibrium.equality.2}), we have
\be
R_{s}u_{s}^*+[(M_s-N_s-\Ga_s)X_s^*+\Phi_s]B_s+M_sD_s'[C_sX_s^*+D_s(\alpha_sX_s^*+\beta_s)+\sg_s]=0,
\;s\in[0,T]
\ee
and hence
\bea
\alpha_s&=&-(R_s+M_sD_s'D_s)^{-1}[(M_s-N_s-\Ga_s)B_s+M_sD_s'C_s],
\\
\beta_s&=&-(R_s+M_sD_s'D_s)^{-1}(\Phi_sB_s+M_sD_s'\sg_s).
\eea

Next, comparing the $ds$ term of $dp_i(s;t)$ in (\ref{FBSDEs.n=1}) and (\ref{dpi.Ansatz})
(we supress the argument $s$ here), we have
\bea
&&M_i[AX^*+B'(\alpha X^*+\beta)+b]+m_iX^*-N_i\{A\mathbb{E}_t[X^*]+B'\mathbb{E}_t[\alpha X^*+\beta]+b\}
-n_i\mathbb{E}_t[X^*]-\ga_iX_t^*+\phi_i
\nonumber\\
&&\qquad=-[A(M_iX^*-N_i\mathbb{E}_t[X^*]-\Ga_iX_t^*+\Phi_i)+M_iC'(CX^*+D(\alpha X^*+\beta)+\sg)].
\eea
Notice in the above that $X^*=X_s^*$ and $\mathbb{E}_t[X^*]=\mathbb{E}_t[X_s^*]$ due to the omission of $s$.
This leads to the following equations for $M_i,N_i,\Ga_i,\Phi_i$:
\bea
&&\left\{
\begin{array}{l}
    \dot{M}_i=-(2A+|C|^2)M_i-Q_i+M_i(B'+C'D)(R+MD'D)^{-1}[(M-N-\Ga)B+MD'C],\;s\in[0,T],
    \\
    M_{i,T}=G_i;
    \end{array}
\right.\label{Eq.of.Mi}
\\
&&\left\{
\begin{array}{l}
    \dot{N}_i=-2AN_i+N_iB'(R+MD'D)^{-1}[(M-N-\Ga)B+MD'C],\;s\in[0,T],
    \\
    N_{i,T}=h_i;
    \end{array}
\right.\label{Eq.of.Ni}
\\
&&\left\{
\begin{array}{l}
    \dot{\Ga}_i=-A\Ga_i,\;s\in[0,T],
    \\
    \Ga_{i,T}=\lambda_i;
    \end{array}
\right.\label{Eq.of.Gai}
\\
&&\left\{
\begin{array}{l}
    \dot{\Phi}_i=-\{A-[B'(M-N)+C'DM](R+MD'D)^{-1}B\}\Phi_i-(M_i-N_i)b-M_iC'\sg
    \\
    \qquad\qquad-[(M_i-N_i)B'+M_iC'D](R+MD'D)^{-1}MD'\sg,\;s\in[0,T],
    \\
    \Phi_{i,T}=-\mu_i.
    \end{array}
\right.\label{Eq.of.Phii}
\eea
Though $M_i,N_i,\Ga_i,\Phi_i,i=1,2$ are scalars,
$M,N,\Ga,\Phi$ are now matrices because of two players.
Therefore, the above equations are more complicated than the similar equations (4.5)-(4.8) in \cite{HJZ}.
Before we solve the equations (\ref{Eq.of.Mi})-(\ref{Eq.of.Phii}),
we first prove that, if exist, the equilibrium constructed above is the unique equilibrium.
Indeed, we have
\begin{theorem}
Let
\be
\mathcal{L}_1=\Bigg\{X(\cdot;\cdot):X(\cdot;t)\in L_\mathcal{F}^2(t,T;\mathbb{R}),
                    \left.\sup_{t\in[0,T]}\mathbb{E}\left[\sup_{s\ge t}|X(s;t)|^2\right]<+\infty\right\}
\ee
and
\be
\mathcal{L}_2=\Bigg\{Y(\cdot;\cdot):Y(\cdot;t)\in L_\mathcal{F}^2(t,T;\mathbb{R}^d),
                    \left.\sup_{t\in[0,T]}\mathbb{E}\left[\int_t^T|X(s;t)|^2ds\right]<+\infty\right\}.
\ee
Suppose all the parameters $A,B_1,B_2,C,D_1,D_2,b,\sg,Q_1,Q_2,R_1$
and $R_2$ are all deterministic.\\
When $(M_i,N_i,\Ga_i,\Phi_i),i=1,2$ exist, and for $i=1,2$, $(p_i(s;t),k_i(s;t))\in\mathcal{L}_1\times \mathcal{L}_2$,
the equilibrium strategy is unique.
\end{theorem}

{\bf Proof. }Suppose there is another equilibrium $(X,u_1,u_2)$, then the equation system (\ref{BSDE.pk}),
with $X^*$ replaced by $X$, admits a solution $(p_i(s;t),k_i(s),u_{i,s})$ for $i=1,2$,
which satisfies $B_{i,s}p_i(s;s)+D_{i,s}'k_i(s)+R_{i,s}u_{i,s}=0$ for a.e. $s\in[0,T]$.
For $i=1,2$, define
\bea
\bar{p}_i(s;t)&\triangleq& p_i(s;t)-[M_{i,s}X_s-N_{i,s}\mathbb{E}_t[X_s]-\Ga_{i,s}+\Phi_{i,s}], \lb{bar.pi}
\\
\bar{k}_i(s;t)&\triangleq& k_i(s)-M_{i,s}(C_sX_s+D_{1,s}u_{1,s}+D_{2,s}u_{2,s}+\sg_s), \lb{bar.ki}
\eea
where $k_i(s)=k_i(s;t)$ by Lemma \ref{Lm:kp.simplify}.

We define $p(s;t)=\diag(p_1(s;t)I_l,p_2(s;t)I_l)$,
$\bar{p}(s;t)=\diag(\bar{p}_1(s;t)I_l,\bar{p}_2(s;t)I_l)$,
and $u=\left(\matrix{u_{1,s}\cr u_{2,s}}\right)$.
By the equilibrium condition (\ref{new.condition}), we have
\bea
0&=&\left(\matrix{B_{1,s}p_1(s;s)+D_{1,s}'k_1(s)+R_{1,s}u_{1,s}\cr
                  B_{2,s}p_2(s;s)+D_{2,s}'k_2(s)+R_{2,s}u_{2,s}}\right)
\nonumber\\
&=&p(s;s)B_s+\left(\matrix{D_{1,s}'k_1(s)\cr D_{2,s}'k_2(s)}\right)+R_su_s
\nonumber\\
&=&[\bar{p}(s;s)+X_s(M_s-N_s-\Ga_s)+\Phi_s]B_s+\left(\matrix{D_{1,s}'\bar{k}_1(s)\cr D_{2,s}'\bar{k}_2(s)}\right)
+M_sD_s'(C_sX_s+D_su_s+\sg_s)+R_su_s
\nonumber\\
&=&\bar{p}(s;s)B_s+\left(\matrix{D_{1,s}'\bar{k}_1(s)\cr D_{2,s}'\bar{k}_2(s)}\right)
+X_s[(M_s-N_s-\Ga_s)B_s+M_sD_s'C_s]+\Phi_sB_s+M_sD_s'\sg_s
\nonumber\\
&&+(R_s+M_sD_s'D_s)u_s.
\eea
Since $R_s+M_sD_s'D_s$ is invertible, we have
\be\label{another.u}
u_s=-(R_s+M_sD_s'D_s)^{-1}\left\{\bar{p}(s;s)B_s+\left(\matrix{D_{1,s}'\bar{k}_1(s)\cr D_{2,s}'\bar{k}_2(s)}\right)
+X_s[(M_s-N_s-\Ga_s)B_s+M_sD_s'C_s]+\Phi_sB_s+M_sD_s'\sg_s\right\},
\ee
and hence for $i=1,2$,
\bea
d\bar{p}_i(s;t)&=&dp_i(s;t)-d[M_{i,s}X_s-N_{i,s}\mathbb{E}_t[X_s]-\Ga_{i,s}+\Phi_{i,s}]
\nonumber\\
&=&-[A_sp_i(s;t)+C_{s}'k_i(s)+Q_{i,s}X_s]ds+k_i'(s)dW_s-d[M_{i,s}X_s-N_{i,s}\mathbb{E}_t[X_s]-\Ga_{i,s}X_t+\Phi_{i,s}]
\nonumber\\
&=&-\bigg\{A_s\bar{p}_i(s;t)+C_s'\bar{k}_i(s)+A_s(M_{i,s}X_s-N_{i,s}\mathbb{E}_t[X_s]-\Ga_{i,s}X_t+\Phi_{i,s})
\nonumber\\
&&\quad+C_s'M_{i,s}(C_sX_s+D_{1,s}u_{1,s}+D_{2,s}u_{2,s}+\sg_s)\bigg\}ds
\nonumber\\
&&\quad+[\bar{k}_i(s)-M_{i,s}(C_sX_s+D_{1,s}u_{1,s}+D_{2,s}u_{2,s}+\sg_s)]'dW_s
\nonumber\\
&&-\bigg\{M_{i,s}[A_sX_s+B_s'u_s+b_s]+m_{i,s}X_s-N_{i,s}(A_s\mathbb{E}_t[X_s]+B_s'\mathbb{E}_t[u_s]+b_s)
\nonumber\\
&&\quad
-n_{i,s}\mathbb{E}_t[X_s]-\ga_{i,s}X_t+\phi_{i,s}\bigg\}ds
\nonumber\\
&&-M_{i,s}[C_sX_s+D_su_s+\sg_s]'dW_s
\nonumber\\
&=&-\Bigg\{A_s\bar{p}_i(s;t)+C_s'\bar{k}_i(s)-M_{i,s}(B_s'+C_s'D_s)(R_s+M_sD_s'D_s)^{-1}\left[B_s\bar{p}(s;s)+\left(\matrix{D_{1,s}'\bar{k}_1(s)\cr D_{2,s}'\bar{k}_2(s)}\right)\right]
\nonumber\\
&&\qquad N_{i,s}B_s'(R_s+M_sD_s'D_s)^{-1}\mathbb{E}_t\left[B_s\bar{p}(s;s)+\left(\matrix{D_{1,s}'\bar{k}_1(s)\cr D_{2,s}'\bar{k}_2(s)}\right)\right]\Bigg\}ds+\bar{k}_i(s)'dW_s,
\eea
where we suppress the subscript $s$ for the parameters,
and we have used the equations (\ref{Eq.of.Mi})-(\ref{Eq.of.Phii}) for $M_i,N_i,\Ga_i,\Phi_i$ in the last equality.
From (\ref{bar.pi}) and (\ref{bar.ki}), we have $(\bar{p}_i,\bar{k}_i)\in\mathcal{L}_1\times\mathcal{L}_2$.
Therefore, by Theorem 4.2 of \cite{HJZ2},
we have $\bar{p}(s;t)\equiv0$ and $\bar{k}(s)\equiv0$.

Finally, plugging $\bar{p}\equiv\bar{k}\equiv0$ into $u$ of (\ref{another.u}),
we get the $u$ being the same form of feedback strategy as in (\ref{u.of.alpha.beta}),
and hence $(X,u_1,u_2)$ is the same as $(X^*,u_1^*,u_2^*)$ which we got before.
\hb

\subsection{Existence of the equilibrium strategies}

The solutions to (\ref{Eq.of.Gai}) is
\be
\Ga_{i,s}=\lambda_ie^{\int_s^T A_tdt},
\ee
for $i=1,2$.
Let $\tilde{N}=N_1\slash N_2$, from (\ref{Eq.of.Ni}), we have $\dot{\tilde{N}}=0$,
and hence
\be
\tilde{N}\equiv {h_1\over h_2},
\quad
N_2\equiv {h_2\over h_1}N_1.
\ee
Equations (\ref{Eq.of.Mi}) and (\ref{Eq.of.Ni}) form a system of coupled Riccati-type equations
for $(M_1,M_2,N_1)$:
\be
\left\{
\begin{array}{l}
    \dot{M}_1=-[2A+|C|^2+B'\Ga(R+MD'D)^{-1}(B+D'C)]M_1-Q_1
    \\
    \qquad\qquad+(B+D'C)'(R+MD'D)^{-1}M(B+D'C)M_1-B'N(R+MD'D)^{-1}(B+D'C)M_1,
    \\
    M_{1,T}=G_1;
    \\
    \dot{M}_2=-[2A+|C|^2+B'\Ga(R+MD'D)^{-1}(B+D'C)]M_2-Q_2
    \\
    \qquad\qquad+(B+D'C)'(R+MD'D)^{-1}M(B+D'C)M_2-B'N(R+MD'D)^{-1}(B+D'C)M_2,
    \\
    M_{2,T}=G_2;
    \\
    \dot{N}_1=-2AN_i+N_iB'(R+MD'D)^{-1}[(M-N-\Ga)B+MD'C],
    \\
    N_{1,T}=h_1.
    \end{array}
\right.\label{Eq.of.M12.N1}
\ee
Finally, once we get the solution for $(M_1,M_2,N_1)$,
(\ref{Eq.of.Phii}) is a simple ODE.
Therefore, it is crucial to solve (\ref{Eq.of.M12.N1}).

Formally, we define $\tilde{M}={M_1\over M_2}$ and $J_1={M_1\over N_1}$
and study the following equation for $(M_1,\tilde{M},J_1)$:
\be
\left\{
\begin{array}{l}
    \dot{M}_1=-[2A+|C|^2+B'\Ga(R+MD'D)^{-1}(B+D'C)]M_1-Q_1
    \\
    \qquad\qquad+(B+D'C)'(R+MD'D)^{-1}M(B+D'C)M_1-B'N(R+MD'D)^{-1}(B+D'C)M_1,
    \\
    M_{1,T}=G_1;
    \\
    \dot{\tilde{M}}=-({Q_1\over M_1}-{Q_2\over M_1}\tilde{M})\tilde{M},
    \\
    \tilde{M}_{T}={G_1\over G_2};
    \\
    \dot{J}_1=-[|C|^2-C'D(R+MD'D)^{-1}M(B+D'C)+B'\Ga(R+MD'D)^{-1}D'C+{Q_1\over M_1}]J_1
    \\
    \qquad\qquad-C'D(R+MD'D)^{-1}M\;\diag(I_l,{h_2\over h_1}\tilde{M}I_l)B,
    \\
    J_{1,T}={G_1\over h_1},
    \end{array}
\right.\label{Eq.of.M1.tildeM.J1}
\ee
where $M=\diag(M_1I_l,{M_1\over\tilde M}I_l), N=\diag({M_1\over J_1}I_l,{h_2\over h_1}{M_1\over J_1}I_l)$
and $\Ga=\diag(\lambda_1e^{\int_s^T A_tdt}I_l,\lambda_2e^{\int_s^T A_tdt}I_l)$.

By a direct calculation, we have
\begin{proposition}
If the system (\ref{Eq.of.M1.tildeM.J1}) admits a positive solution $(M_1,\tilde{M},J_1)$,
then the system (\ref{Eq.of.M12.N1}) admits a solution $(M_1,M_2,N_1)$.
\end{proposition}

In the following, we will use the truncation method to study the system (\ref{Eq.of.M1.tildeM.J1}).
For convenienc, we use the following notations:
\bea
a\vee b &=&\max\{a,b\},\qquad \forall a,b\in\mathbb{R},
\\
a\wedge b&=& \min\{a,b\},\qquad \forall a,b\in\mathbb{R}.
\eea
Moreover, for a matrix $M\in\mathbb{R}^{m\times n}$ and a real number $c$, we define
\bea
(M\vee c)_{i,j}=M_{i,j}\vee c,\qquad \forall 1\le i\le m,1\le j\le n,
\\
(M\wedge c)_{i,j}=M_{i,j}\wedge c,\qquad \forall 1\le i\le m,1\le j\le n.
\eea

We first consider the standard case where $R-\delta{I}\succeq0$ for some $\delta>0$.
We have
\begin{theorem}
Assume that $R-\delta{I}\succeq0$ for some $\delta>0$ and $G\ge h>0$.
Then (\ref{Eq.of.M1.tildeM.J1}), and hence (\ref{Eq.of.M12.N1}) admit unique solution if

(i) there exists a constant $\lambda\ge0$ such that $B=\lambda D'C$;

(ii) $\frac{|C|^2}{2l}D'D-(\lambda+1)D'CC'D\succeq0$.
\end{theorem}

{\bf Proof.}
For fixed $c>0$ and $K>0$, consider the following truncated system of (\ref{Eq.of.M1.tildeM.J1}):
\be
\left\{
\begin{array}{l}
    \dot{M}_1=-[2A+|C|^2+B'\Ga(R+M_c^+D'D)^{-1}(B+D'C)]M_1-Q_1
    \\
    \qquad\qquad+(B+D'C)'(R+M_c^+D'D)^{-1}(M_c^+\wedge K)(B+D'C)M_1-B'(N_c^+\wedge K)(R+M_c^+D'D)^{-1}(B+D'C)M_1,
    \\
    M_{1,T}=G_1;
    \\
    \dot{\tilde{M}}=-({Q_1\over M_1\vee c}-{Q_2\over M_1\vee c}\tilde{M}\wedge K)\tilde{M},
    \\
    \tilde{M}_{T}={G_1\over G_2};
    \\
    \dot{J}_1=-\lambda^{(1)}J_1-C'D(R+M_c^+D'D)^{-1}(M_c^+\wedge K)\diag(I_l,{h_2\over h_1}(\tilde{M}\wedge K)I_l)B,
    \\
    J_{1,T}={G_1\over h_1},
    \end{array}
\right.\label{truncated.Eq.of.M1.tildeM.J1}
\ee
where $M_{c}^+=\diag((M_1\vee0)I_l,{{M_1\vee0}\over\tilde M\vee c}I_l)$,
$N_c^+=\diag({{M_1\vee0}\over J_1\vee c}I_l,{h_2\over h_1}{{M_1\vee0}\over J_1\vee c}I_l)$
and
\be
\lambda^{(1)}=|C|^2-C'D(R+M_c^+D'D)^{-1}(M_c^+\wedge K)(B+D'C)+B'\Ga(R+M_c^+D'D)^{-1}D'C+{Q_1\over M_1\vee c}.
\ee

Since $R-\delta I\succeq0$,
the above system (\ref{truncated.Eq.of.M1.tildeM.J1}) is locally Lipschitz with linear growth,
and hence it admits a unique solution $(M_1^{c,K},\tilde{M}^{c,K},J_1^{c,K})$.
We will omit the superscript $(c,K)$ when there is no confusion.

We are going to prove that $J_1\ge1$ and that $M_1,\tilde{M}\in[L_1,L_2]$
for some $L_1,L_2>0$ independent of $c$ and $K$ appearing in the truncation functions.
We denote
\bea
\lambda^{(2)}&=&(2A+|C|^2+B'\Ga(R+M_c^+D'D)^{-1}(B+D'C))
\nonumber\\
&&\qquad
-(B+D'C)'(R+M_c^+D'D)^{-1}(M_c^+\wedge K)(B+D'C)
\nonumber\\
&&\qquad-B'(N_c^+\wedge K)(R+M_c^+D'D)^{-1}(B+D'C).
\eea
Then $\lambda^{(2)}$ is bounded, and $M_1$ satisfies
\be
\dot{M}_1+\lambda^{(2)}M_1+Q_1=0,\quad M_{1,T}=G_1.
\ee
Hence $M_1>0$.
Similarly, we have $\tilde{M}>0$.

The equation for $\tilde{M}$ is
\be
\left\{
\begin{array}{l}
    -\dot{\tilde{M}}=({Q_1\over M_1\vee c}\tilde{M}-{Q_2\over M_1\vee c}(\tilde{M}\wedge K)\tilde{M},
    \\
    \tilde{M}_{T}={G_1\over G_2};
    \end{array}
\right.\label{truncated.Eq.of.tildeM}
\ee
hence $\tilde{M}$ admits  an upper bound $L_2$ independent of $c$ and $K$.
Choosing $K=L_2$ and examining again (\ref{truncated.Eq.of.tildeM}),
we deduce that there exists $L_1>0$ independent of $c$ and $K$ such that $\tilde{M}\ge L_1$.
Indeed, we can choose $L_1=\min_{0\le t\le T}{Q_{1,t}\over Q_{2,t}}\wedge{G1\over G_2}$
and $L_2=\max_{0\le t\le T}{Q_{1,t}\over Q_{2,t}}\vee{G1\over G_2}$.
As a result, choosing $c<L_1$, the terms $M_c^+$ can be replaced by $M=\diag(M_1I_l,{M_1\over\tilde M}I_l)$,
respectively, in (\ref{truncated.Eq.of.M1.tildeM.J1}) without changing their values.

Now we prove $J\ge1$. Denote $\tilde{J}=J_1-1$, then $\tilde{J}$ satisfies the ODE:
\be
\dot{\tilde{J}}=-\lambda^{(1)}\tilde{J}-[\lambda^{(1)}+C'D(R+MD'D)^{-1}(M\wedge K)\diag(I_l,{h_2\over h_1}\tilde{M}I_l)B]=-\lambda^{(1)}\tilde{J}-a^{(1)},
\ee
where
\bea
a^{(1)}&=&\lambda^{(1)}+C'D(R+MD'D)^{-1}(M\wedge K)\diag(I_l,{h_2\over h_1}\tilde{M}I_l)B
\nonumber\\
&=&|C|^2-(\lambda+1)C'D(R+MD'D)^{-1}(M\wedge K)D'C+C'D\Ga(R+MD'D)^{-1}(M\wedge K)D'C++{Q_1\over M_1\vee c}
\nonumber\\
&&+C'D(R+MD'D)^{-1}(M\wedge K)\diag(I_l,{h_2\over h_1}\tilde{M}I_l)D'C
\nonumber\\
&&\ge|C|^2-(\lambda+1)C'D(R+MD'D)^{-1}MD'C+C'D\Ga(R+MD'D)^{-1}(M\wedge K)D'C++{Q_1\over M_1\vee c}
\nonumber\\
&&=tr\left\{(R+MD'D)^{-1}{|C|^2+Q_1\slash(M_1\vee c)\over2l}(R+MD'D)\right\}-(\lambda+1)tr\{(R+MD'D)^{-1}D'CC'DM\}
\nonumber\\
&&=tr\left\{(R+MD'D)^{-1}H\right\}
\eea
with $H={|C|^2+Q_1\slash(M_1\vee c)\over2l}(R+D'DM)-(\lambda+1)D'CC'DM$.

When $c$ is small enough such that $R-cD'D\succeq0$, we have
\be
{Q_1\over M_1\vee c}(R+MD'D)\ge {Q_1\over L_2}D'D.
\ee
Hence,
\be
H\succeq({|C|^2\over2l}D'D-(\lambda+1)D'CC'D)M\succeq0,
\ee
and consequently $a^{(1)}\ge tr\{(R+MD'D)^{-1}H\}\ge0$.
We then deduce that $\tilde{J}\ge0$, and hence $J_1\ge1$.
The boundness of $M_1$ can be proved by a similar argument in the proof of Theorem 4.2 in \cite{HJZ}.
\hb

Similarly, for the singular case $R\equiv0$, we have
\begin{theorem}
Given $G_1\ge h_1\ge1,R\equiv0$,
if $B=\lambda D'C$ and $|C|^2-(\lambda+1)C'D(D'D)^{-1}D'C\ge0$,
then (\ref{Eq.of.M1.tildeM.J1}) and (\ref{Eq.of.M12.N1}) admit a unique positive solution.
\end{theorem}

Concluding the above two theorems,
we can present our main results of this section:
\begin{theorem}
Given $G_1\ge h_1\ge1$ and $B=\lambda D'C$.
The (\ref{Eq.of.M12.N1}) admits a unique positive solution $(M_1,M_2,N_1)$ in the following two cases:

(i) $R-\delta I\succeq0$ for some $\delta>0$, $\frac{|C|^2}{2l}D'D-(\lambda+1)D'CC'D\succeq0$;

(ii) $R\equiv0$, $|C|^2-(\lambda+1)C'D(D'D)^{-1}D'C\ge0$.
\end{theorem}

{\bf Proof.}
Define $p_i(s;t)$ and $k_i(s;t)$ by (\ref{Ansatz}) and (\ref{ki.Ansatz}), respectively.
It is straightforward to check that $(u_1^*,u_2^*,X^*,p_1,p_2,k_1,k_2)$ satisfies the system of SDEs (\ref{FBSDEs.n=1}).
Moreover, in the both cases, we can check that $\alpha_{i,s}$ and $\beta_{i,s}$ in (\ref{u.of.alpha.beta}) are all uniformly bounded,
and hence $u_i^*\in L_{\mathcal{F}}^2(0,T;\mathbb{R}^l)$ and $X^*\in L^2(\Omega;C(0,T;\mathbb{R}))$.

Finally, denote $\Lambda_i(s;t)=R_{i,s}u_{i,s}^*+p_i(s;t)B_{i,s}+(D_{i,s})'k_i(s;t),i=1,2$.
Plugging $p_i,k_i,u_i^*$ define in (\ref{Ansatz}),(\ref{ki.Ansatz}) and (\ref{u.of.alpha.beta}) into $\Lambda_i$,
we have
\be
\Lambda_i(s;t)=R_{i,s}u_{i,s}^*+(M_{i,s}X_s^*-N_{i,s}\mathbb{E}_t[X_s^*]-\Ga_{i,s}X_t^*+\Phi_{i,s})B_{i,s}
+M_{i,s}D_{i,s}'[C_sX_s^*+D_{1,s}u_{1,s}^*+D_{2,s}u_{2,s}^*+\sg_s]
\ee
and hence,
\bea
\Lambda(t;t)&\triangleq&\left(\matrix{\Lambda_1(t;t)\cr \Lambda_2(t;t)}\right)
\nonumber
\\
&=&(R_t+M_tD_t'D_t)u_t^*+M_t(B_t+D_t'C_t)X_t^*
-N_tB_t\mathbb{E}_t[X_t^*]-\Ga_tB_tX_t^*+(\Phi_tB_t+M_tD_t'\sg_t)
\nonumber\\
&=&-[(M_t-N_t-\Ga_t)B_t+M_tD_t'C_t]X_t^*-(\Phi_tB_t+M_tD_t'\sg_t)
\nonumber\\
&&\quad+M_t(B_t+D_t'C_t)X_t^*-N_tB_tX_t^*-\Ga_tB_tX_t^*+(\Phi_tB_t+M_tD_t'\sg_t)
\nonumber\\
&=&0.
\eea
Therefore, $\Lambda_i$ satisfies the seond condition in (\ref{new.condition}).
\hb

%

\medskip


\bibliographystyle{abbrv}

\begin{thebibliography}{}

\bibitem{BaC}S. Basak, G.Chabakauri, Dynamic mean-variance asset allocation.
{\it The Review of Financial Studies}. (2010)23:2970-3016.

\bibitem{Ben}A. Bensoussan, K.C.J. Sung, S.C.P. Yam. Linear-Quadratic Time-Inconsistent Mean Field Games.
{\it Dynamic Games and Applications.} (2013) 3:537-552.

\bibitem{BjM} T. Bj\"{o}rk, A. Murgoci, A general theorey of Markovian time inconsistent stochastic control problem.
1694759, Social Science Research Network(SSRN). {\it http://papers.ssrn.com/so13/papers.cfm?abstract\_id=1694759}. (2010).

\bibitem{BMZ} T. Bj\"{o}rk, A. Murgoci, X.Zhou, Mean-variance portfolio optimization with state-dependent risk aversion.
 {\it Mathematical Finance}. (2014)24:1-24.

\bibitem{Eke1}I. Ekeland, From Ramsey to Thom: a classical problem in the calculus of variations leading to an implicit differential equation, {\it Discrete and Continuous Dynamical Systems}. (2010)28:1101 - 1119.

\bibitem {EkL}I. Ekeland, A. Lazrak, The golden rule when preferences are time-inconsistent.
{\it Mathematics and Financial Economics}. (2010)4:29-55.

\bibitem{EkL1}I. Ekeland, A. Lazrak, Equlibrium policies when preferences are time-inconsistent, {\it http://arxiv.org/abs/math/0808.3790}.(2006).

\bibitem{EkL2}I. Ekeland, A. Lazrak, Being serious about non-commitment: subgame perfect equilibrium in continuous time, {\it http://arxiv.org/abs/math/0604264}.(2006).

\bibitem {EKS}I. Ekeland, L. Karp, R. Sumaila,Equilibrium management of fisheries
with overlapping altruistic generations.
{\it http://www.ceremade.dauphine.fr/\symbol{126}ekeland/Articles/Karp.pdf}

\bibitem {ELZ}I. Ekeland, Y. Long, Q. Zhou, A new class of problems in the calculus of variations.
{\it Regular and Chaotic Dynamics.} (2013)258:553-584.

\bibitem{EkP}I. Ekeland, T.A. Pirvu, Investment and consumption without commitment,
    {\it Mathematical anf Financial Eonomics}. (2008)2:57-86.

\bibitem{GrW}S.R. Grendadier, N. Wang, Investment under uncertianty and time-inconsistent preferences.
{\it Journal of Financial Economics.} (2008)2:57-86.

\bibitem{HJZ} Y. Hu, H. Jin, X. Zhou, Time-Inconsistent Stochastic Linear-quadratic Control. {\it SIAM Journal of Control and Optimization.} (2012)50:1548-1572.

\bibitem{HJZ2} Y. Hu, H. Jin, X. Zhou, Time-Inconsistent Stochastic Linear--Quadratic Control: Characterization and Uniqueness of Equilibrium. {\it http://arxiv.org/pdf/1504.01152.pdf.} (2015).

\bibitem {Kar2}L. Karp, Non-constant discounting in continuous time.
{\it Journal of Economic Theory}. (2007)132:577-568.

\bibitem {KaL1}L. Karp, I. H. Lee, Time-consistent policies.
{\it Journal of Economic Theory}. (2003)112:353-364.

\bibitem {Krusell}P. Krusell, A. Smith, Consumption-savings decisions with quasi-geometric discounting.
{\it Econometrica}. (2003)71:365--375.

\bibitem {Laibson}C. Harris, D.Laibson, Dynamic choices of hyperbolic consumers.
{\it Econometrica}. (2001)69:935--957.

\bibitem{Mar}J. Mart\'in-Solano, Group inefficiency in a common property resource game with asymmetric players.
{\it http://papers.ssrn.com/sol3/papers.cfm?abstract\_id=2516846.} (2014).

\bibitem{Peng}S. Peng, A general stochastic maximm principle for optimal control problems.
{\it SIAM Journal of Control and Optimization.} (1990)28:966-979.

\bibitem {Phelps}E. Phelps, R. A. Pollak, On second-best national saving and game-equilibrium growth.
{\it The Review of Economic Stududies}. (1968)35:185-199.

\bibitem {Phelps1}E. Phelps, The indeterminacy of game-equilibrium growth.
In: Phelps, E.S. (ed.) "Altruism, Morality and Economic theory", pp.
87--105. {\it Russell Sage Foundation, New York}. (1975).

\bibitem{Stroz}R.H.Stroz, Myopia and inconsistency in dynamic utility maximization.
{\it The Review of Economic Studies.} (1955) 165-180.

\bibitem{Yon} J. Yong, A deterministic linear quadratic time-inconsistent optimal control problems.
{\it Mathematical Control and Related Fields}. (2011) 83-118.

\bibitem{YoZ} J. Yong, X. Zhou, Stochastic controls: Hamiltonian Systems and HJB Equations, {\it Springer-Verlag, NewYork}. (1999).
\end{thebibliography}

\end{document}